\documentclass[aip,jap,amsmath,amssymb,
reprint
]{revtex4-1}

\usepackage{graphicx}
\usepackage{hyperref}

\makeatletter
\def\boldit#1{{\reset@font\bfseries\itshape#1}}
\makeatother

\begin{document}





\title{\boldit{Ab inito} calculation of the local magnetic moment in titanium doped zinc oxide with a corrected-band-gap scheme} 








\author{Bin Shao}





\affiliation{College of Information Technical Science, Nankai University, Tianjin 300071, China.}

\author{Hong Liu}





\affiliation{Office of International Academic Exchanges, Nankai University, Tianjin 300071, China.}

\author{Jian Wu}





\affiliation{Physics Department, Tsinghua University, Beijing 100084, China.}





\author{Xu Zuo}





\affiliation{College of Information Technical Science, Nankai University, Tianjin 300071, China.}

\date{\today}

\begin{abstract}


The local magnetic moment of Ti:ZnO is calculated from first principles by using the corrected-band-gap scheme (CBGS). The result shows that the system is magnetic with the magnetization of 0.699 $\mu_\text{B}$ per dopant. The origin of the local magnetic moment is considered to be the impurity band partially occupied by the donor electrons in the conduction band. Further, the impacts of applying Hubbard \textit{U} to Ti-\textit{d} orbital on the magnetic moment have been investigated.

\end{abstract}



\maketitle 




Zinc oxide, a diamagnetic semiconductor, doped by certain transition metals (TM) was observed to be a room-temperature ferromagnet. Moreover, according to the experiment of Venkatesan \textit{et al.},\cite{JMDCoey} \textquotedblleft there is a striking systematic variation (of magnetization), with maxima near the beginning of the 3\textit{d} series at Ti and V (0.5 $\mu_\text{B}$)...\textquotedblright .

To understand the origin of the magnetic moment, first-principle calculations for TM doped ZnO have been carried out, especially for cobalt. However, there are few reports on Ti doped ZnO. Early calculation \cite{Chien2004} using density functional theory (DFT) approach based on GGA (Generalized Gradient Approximation) and GGA+\textit{U} (GGA plus on-site Coulomb repulsion) predicted a magnetic moment (0.9 $\mu_\text{B}$ and 1.6 $\mu_\text{B}$, respectively) larger than the experimental result\cite{JMDCoey}. In addition, structure optimization was ignored in Ref.~\onlinecite{Chien2004}. Further work \cite{Osuch} with atomic-position relaxation using full potential linear augmented plane wave method obtains a result of 0.63 $\mu_\text{B}$ per dopant, which is in good accordance with the experiment.\cite{JMDCoey} However, both of the previous works may be affected by the well-known problem of local-spin-density approximation (LSDA)/GGA, underestimation of the band gap. As a result, they are suffered from the incorrect impurity levels relative to the host band edges, which may significantly impact the calcualted local magnetic moment originating from the impurity levels. In this work, we investigate Ti:ZnO using a corrected-band-gap scheme (CBGS), where both band gap and impurity levels are corrected by the DFT+\textit{U} approach\cite{Dudarev}.


A $3\times{3}\times{2}$ supercell with 72 atoms is constructed from the wurtzite primitive cell of ZnO with the experimental lattice parameters, and one Zn atom is substituted by Ti. The atomic positions are first optimized by using Perdew-Burke-Ernzerhof (PBE) \cite{PBE} parameterization of GGA and projected augmented wave (PAW) method. The optimization is accomplished by using VASP code\cite{vasp} on a $3\times{3}\times{3}$ \textit{k}-mesh. It shows that the four O atoms nearest neighbor to Ti move inward. As shown in Table~\ref{TableI}, the lengths of the four Ti-O bonds, one along the crystallographic \textit{c} axis (the unique) and others in the basal plane, are all shortened to 1.88 {\AA}, which are originally 1.99 {\AA} and 1.97 {\AA} in the primitive cell of ZnO, respectively. As a result, the local symmetry changes from $C_{3v}$ to $T_{d}$. 

 \begin{table}
 	\caption{\label{TableI} The lengths ({\AA}) of Ti-O bonds. The unique refers to the bond along the \textit{c} axis, and the other refers to those bonds in the basal plane.}
 	\begin{ruledtabular}
		\begin{tabular}{lcc}
 			&unique&other\\ \hline
 			initial&1.99&1.97\\
 			optimization&1.88&1.88\\
 		\end{tabular}
 	\end{ruledtabular}
 \end{table}
 
After structural optimization, the electronic structure and magnetic properties of Ti:ZnO are calculated with the CBGS, by which the Hubbard \textit{U}s applied to Zn-\textit{d} orbital (5 eV) and O-\textit{s} orbital (-50 eV)  enhance the band gap of pristine ZnO to 2.8 eV\cite{WeiS-H}. The \textit{k}-mesh increases to $4\times{4}\times{4}$ and the plane-wave cutoff is 500 eV. Furthermore, with CBGS, calculation with Hubbard \textit{U} applied on Ti-\textit{d} orbitals ($U_{\text{Ti}-d}$) are performed to investigate the impacts of \textit{d}-electron correlation.

\begin{figure}
 \includegraphics{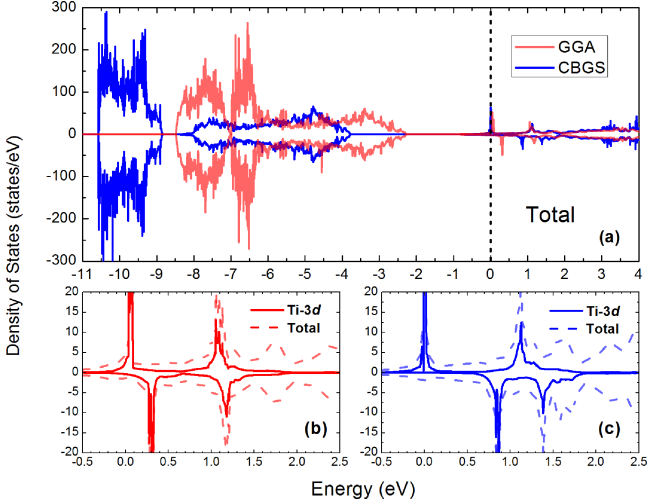}%
 \caption{\label{Fig1}(a) The total DOS by GGA and CBGS, and the projected DOS of Ti-3\textit{d} (b) by GGA and (c) by CBGS.}%
\end{figure}

In Table~\ref{TableII}, we summarize the results of magnetic moments by different methods. Obviously, the magnetic moments are increased by CBGS and further enhanced by CBGS+\textit{U}, and reach the largest value when \textit{U}=3 eV. The total density of states (DOS) [Fig.~\ref{Fig1}(a)] shows that the gap between the O-\textit{p} band and Ti-\textit{d} band is preserved in Ti:ZnO with CBGS. The valence band (VB) of the host and the Zn-\textit{d} band are pushed down to lower energy by CBGS. Moreover, the hybrid between Zn-\textit{d} band and O-\textit{p} band is reduced. These changes due to CBGS are observable in Co:ZnO as well \cite{WeiS-H}. As shown in Figs.~\ref{Fig1} (b) and (c), CBGS increases the exchange split of Ti-\textit{d} bands from 0.25 eV to 0.75 eV. The majority Ti-\textit{d} band lies at the conduction-band minimum (CBM) and straddles the Fermi level. Thus, the system is metallic. 

 \begin{table}
 	\caption{\label{TableII} The magnetic moment given by different methods. }
	\begin{ruledtabular}
 		\begin{tabular}{lccc}
 		&\multicolumn{2}{c}{Magnetic moment ($\mu_\text{B}$)}&\\ \cline{2-3}
 		Method&Total&Ti&$U_{\text{Ti}-d}$ (eV)\\ \hline
 		GGA&0.227&0.199&|\\
 		CBGS&0.699&0.703&|\\
 		CBGS+$U_{\text{Ti-}d}$&0.777&0.809&2.0\\
 		&0.973&0.957&3.0\\
 		\end{tabular}
 	\end{ruledtabular}
 \end{table}

\begin{figure}
 \includegraphics{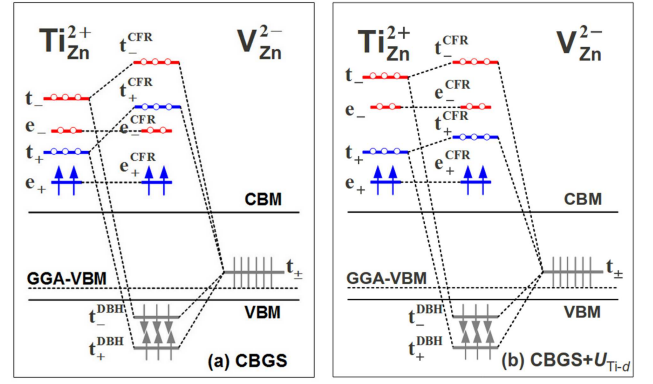}%
 \caption{\label{Fig2}Formation of defect levels in Ti:ZnO with (a) CBGS and (b) CBGS+$U_{\text{Ti-}d}$. The $\text{Ti}^{2+}_{\text{Zn}}$ orbitals in tetrahedral site splits into two-fold \textit{e} and three-fold \textit{t} levels in crystal field. Then the \textit{t} levels interact with the dangling-bond states of the Zn vacancy, and form the defect levels.}%
\end{figure}

To understand the origin of the local magnetic moment, we introduce a model about the formation of the defect levels (Fig.~\ref{Fig2}). According to Zunger \textit{et al.}, \cite{Raebiger,Raebiger2008} the defect levels are a result of the hybridization between the levels of the host cation vacancy (anion dangling bonds) and the 3\textit{d} orbitals of the impurity under the ligand field. As shown in Fig.~\ref{Fig2}, we provide the diagram for Ti substitution in ZnO. Because of the conductive character of Ti-\textit{d} electrons, the defect levels, two hybrid \textit{t} levels and two nonbonding \textit{e} levels, lie above CBM. Since the $e_{+}^{\text{CFR}}$ is occupied by 2\textit{e}, the magnetic moment should be 2 $\mu_\text{B}$ by the first Hund's rule. Fig.~\ref{Fig2} also illustrates the impacts of adding \textit{U} on Ti-\textit{d} on the defect levels, that the exchange split energy is enhanced by applying \textit{U}, and that the minority defect level $e_{-}^{\text{CFR}}$ moves to the energy higher than the majority defect level $t_{+}^{\text{CFR}}$. 

\begin{figure}
 \includegraphics{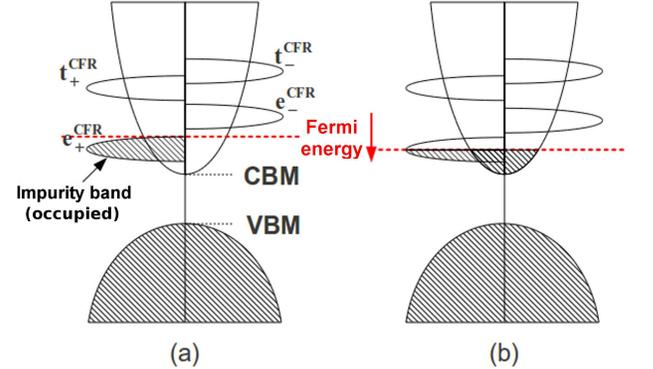}%
 \caption{\label{Fig3}The impacts of band filling effect on magnetic moment. The occupied impurity band is higher than the host CBM. Consequently, part of the donor electrons transfers from the impurity band to the host CBs, and the Fermi energy lowers down.}%
\end{figure}

According to the above analysis (Fig.~\ref{Fig2}), the total magnetic moment should be 2 $\mu_\text{B}$ per dopant. However, these values in Table~\ref{TableII} are smaller than 1 $\mu_\text{B}$. The deviation is caused by the band filling effect. As shown in Fig.~\ref{Fig3}(a), the defect levels form the impurity bands. As the lowest impurity band occupied by two Ti-\textit{d} electrons (donor electrons) are above the host CBM, part of the donor electrons transfers from the impurity band to the host CBs, and the Fermi energy lowers down [Fig.~\ref{Fig3}(b)]. In addition, we integrate the DOS of CBGS (Fig.~\ref{Fig1}) from CBM to Fermi level and obtain the electronic populations of 1.4 and 0.6 for the majority and  minority spins, respectively. Hence, the total number of donor electron is 2.0, supporting the defect level model (Fig.~\ref{Fig2}) and the band filling model (Fig.~\ref{Fig3}). 

\begin{figure}
 \includegraphics{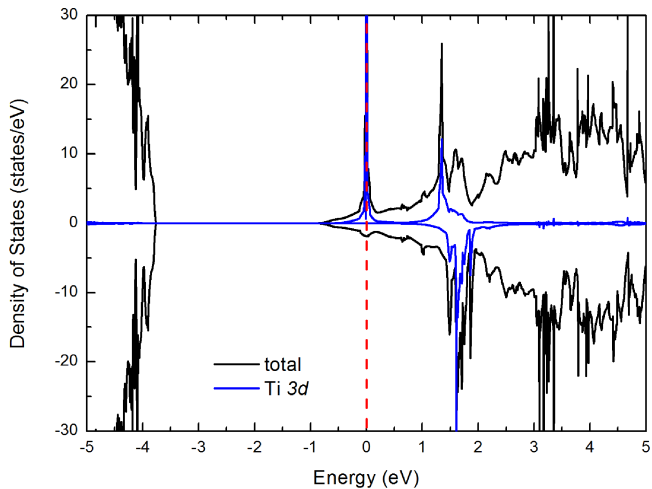}%
 \caption{\label{Fig4}The projected DOS with CBGS+$U_{\text{Ti-}d}$ (2.0 eV).}%
\end{figure}

\begin{figure}
 \includegraphics{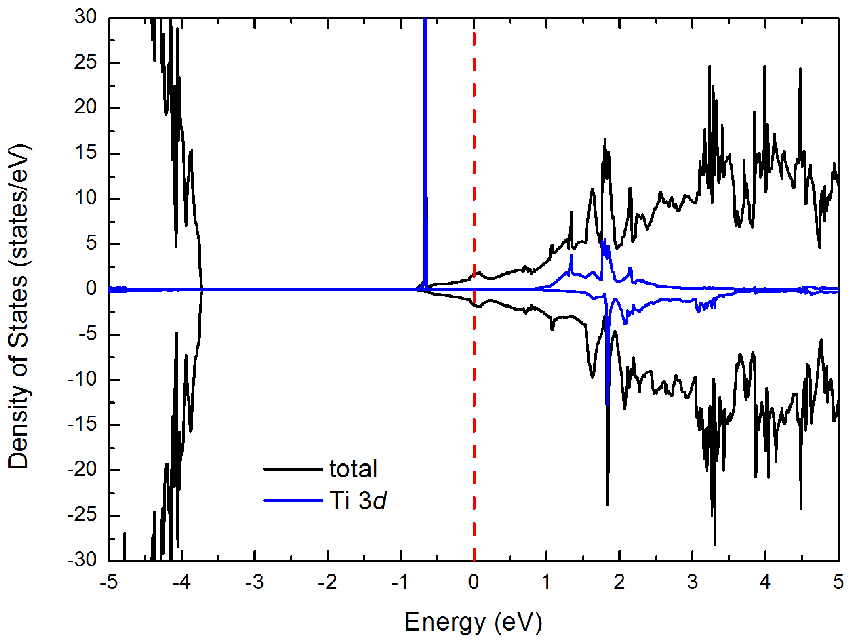}%
 \caption{\label{Fig5}The projected DOS with CBGS+$U_{\text{Ti-}d}$ (3.0 eV).}%
\end{figure}

\begin{figure}
 \includegraphics{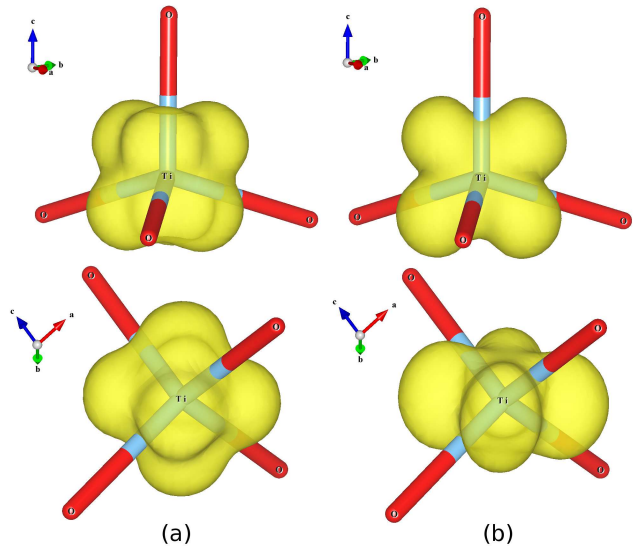}%
 \caption{\label{Fig6}The spin density maps from two respects with CBGS+$U_{\text{Ti-}d}$, (a) \textit{U}=2.0 eV and (b) \textit{U}=3.0 eV. When \textit{U}=2.0 eV, the shape of spin density is obversely a combination of two \textit{d} orbitals, $d_{3z^2-r^2}$ and $d_{x^2-y^2}$. However, that of \textit{U}=3.0 eV is just one of the two \textit{d} orbitals, $d_{x^2-y^2}$.}%
\end{figure}

By applying Hubbard \textit{U} on Ti-3\textit{d} orbital, the exchange split of Ti-\textit{d} band continues to increase and the positions of the impurity bands in different spins support the model in Fig.~\ref{Fig2}(b), as shown in the DOS with CBGS+$U_{\text{Ti-}d}$ (Figs.~\ref{Fig4} and \ref{Fig5}). When \textit{U}=3.0 eV, the double degenerated and the partially occupied impurity band corresponding to the $e_{+}^{\text{CFR}}$ level splits into two bands, one of which is occupied by 1\textit{e} and results in the magnetic moment of about 1 $\mu_\text{B}$ (Table~\ref{TableII}). In general, a well-accepted value of \textit{U} for Ti-3\textit{d} orbitals is 2.0 eV\cite{Raebiger}. At the same time, we also observe the changing of the spin density associated with the split of degenerated \textit{e} orbitals as \textit{U} increases from 2 to 3 eV (Fig.~\ref{Fig6}). When \textit{U}=2.0 eV [Fig.\ref{Fig6}(a)], the shape of spin density is obviously a combination of two \textit{d} orbitals, $d_{3z^2-r^2}$ and $d_{x^2-y^2}$. However, that of \textit{U}=3.0 eV [Fig.\ref{Fig6}(b)] is just one of the two \textit{d} orbitals, $d_{x^2-y^2}$.

In conclusion, we calculate the local magnetic moment of Ti:ZnO from first principles with CBGS. The result shows that the system is magnetic with 0.699 $\mu_\text{B}$ per dopant. The origin of the local magnetic moment is also discussed. We conclude that, due to the position of the defect levels and band filling effect, the Ti-\textit{d} band is partially occupied by donor electrons. Further, we investigate the impacts of adding \textit{U} to Ti-\textit{d} orbitals on the magnetic moment of the system. Both DOS and spin density map indicate that increasing $U_{\text{Ti}-d}$ leads to a further split of Ti-\textit{d} orbitals and simultaneously causes a larger magnetic moment.


\begin{acknowledgments}
This research is sponsored by National Natural Science Foundation of China (Grant No. 10970499), National Basic Research Program of China (973 Program, Grant No. 2011CB606405), and Tianjin Municipal Science and Technology Commission (Grant No. 08JCYBJC12900).
\end{acknowledgments}


%














%




%







%







\end{document}